\documentclass[aps,preprint]{revtex4}%
\usepackage{amsfonts}
\usepackage{amsmath}
\usepackage{amssymb}
\usepackage{graphicx}%
\providecommand{\U}[1]{\protect\rule{.1in}{.1in}}

\begin{document}
\preprint{UATP/0906}
\title[Thermodynamics of relaxation]{Non-equilibrium Thermodynamics:\ Structural Relaxation, Fictive temperature
and Tool-Narayanaswamy phenomenology in Glasses}
\author{P. D. Gujrati}
\affiliation{Departments of Physics and Polymer Science, The University of Akron, OH 44325}

\begin{abstract}
Starting from the second law of thermodynamics applied to an isolated system
consisting of the system surrounded by an extremely large medium, we formulate
a general non-equilibrium thermodynamic description of the system when it is
out of equilibrium. We then apply it to study the structural relaxation in
glasses and establish the phenomenology behind the concept of the fictive
temperature and of the empirical Tool-Narayanaswamy equation on firmer
theoretical foundation.

\end{abstract}
\date{\today}
\maketitle

\section{Introduction}

It is well known that when a liquid is disturbed suddenly from its equilibrium
state by changing the temperature or pressure of the surrounding medium, or
both, then the liquid undergoes a rapid, solidlike change, followed by a
slower, liquidlike change towards the new equilibrium state. These changes can
be seen in the variation in its thermodynamic properties such as the volume
$V$ or its enthalpy $H$ with time. For a supercooled liquid, the above
scenario plays an important role. As the temperature is lowered or the
pressure is increased, the scale separation between the fast and the slow
processes in supercooled liquids increases until the latter becomes too large
compared to the experimental observation time $\tau_{\text{obs}}$. In this
case, the system is said to be kinetically arrested in that the liquidlike
changes no longer contribute to the observed properties. The system behaves
like a solid and is called a \emph{glass}
\cite{GoldsteinSimha,Jones,Dyre,Scherer,Scherer90,Hodge}.

The glass is a system that may be far from equilibrium so one cannot apply
equilibrium statistical mechanics to investigate its properties, which vary
with time. One must resort to apply non-equilibrium thermodynamics
\cite{deGroot,Ottinger}, not a well-established field at present, to study
glasses and their relaxation in time; the latter are usually known as
\emph{structural relaxation}. It is known from the early work of Littleton
\cite{Littleton}, and Lillie \cite{Lillie} that the isothermal viscosity of a
glass changes during relaxation, thus implying the dependence of the
relaxation on changes in the state of the glass \cite{Cox}. The most general
framework for developing non-equilibrium thermodynamics must satisfy the
second law of thermodynamics or must start from it. The generality arises
since the second law is independent of the processes (structural or otherwise)
going inside the system. It is also independent of the details of the system
considered and does not requires any sophisticated concepts like ergodicity or
its loss, etc. Our main goal in this work is to develop an appropriate
non-equilibrium thermodynamics, which will then be applied to glasses with the
hope to gain some new insight and to clarify at the fundamental level certain
concepts extensively used in glasses.

Glassy behavior and their properties have been extensively studied and usually
explained by invoking empirical rules \cite{Tool, Narayanaswamy,Hodge} that,
although they have proved invaluable and very reliable, lack theoretical
justifications \cite{Scherer,Scherer90}. Only recently, attempts have been
made \cite{Wolynes,Schweizer} that use modern framework to investigate
non-equilibrium properties of glassy relaxation at the molecular level. Our
treatment here differs from these attempts in that we develop our approach
using the second law that should be applicable to all systems including
glasses. We do not derive the actual laws of relaxation for which one must
turn to other sources such as \cite{Wolynes,Schweizer}. Our goal is quite
different. We wish to understand some of the important concepts used for
glasses at a fundamental level. One of the most widely used concept in this
field is that of the \emph{fictive temperature}, first introduced by Tool
\cite{Tool} in an empirical fashion to describe non-linear relaxation in
glasses. The system under study slows down so much upon reducing the
temperature from its initial value $T^{\prime}$, where the system was in
equilibrium, to some temperature $T_{0}$\ that one has to wait for a very long
time before true equilibrium is reached at the final temperature. In this
case, crudely speaking, the glass properties are assumed to be similar to a
fictive liquid at some intermediate temperature between $T^{\prime}$ and
$T_{0}$. As time goes on and as the system undergoes structural rearrangements
to come to equilibrium, the temperature of the system continues to change and
finally becomes $T_{0}$. This already means that the fictive temperature of
the system continuously changes from $T^{\prime}$\ to $T_{0}$. Despite its
continual usage in the field, the true meaning of the fictive temperature,
though phenomenologically obvious, is not well-defined in terms of fundamental
quantities such as the entropy. In particular, there exists a variety of
fictive temperatures, each associated with the relaxing quantity under
investigation, which makes the concept not very rich.

There is another aspect of structural relaxation. Its presence means that the
glass is a non-equilibrium state. Thus, its temperature must be changing
during the process of relaxation. How does one define the\emph{ instantaneous
temperature} of the liquid? The instantaneous temperature itself must relax to
$T_{0}$\ as time goes on. Thus, there will a relaxation time describing the
relaxation of the temperature of the glass. Tool \cite{Tool} and Narayanaswamy
\cite{Narayanaswamy}, among others, observed that the relaxation time not only
depends upon the temperature $T_{0}$, but also depends upon the fictive
temperature of the system; see for example \cite{Scherer,Scherer90,Jones}. Is
the instantaneous temperature of the glass the same as the fictive
temperature? These are important issues as a deeper understanding of these
concepts will provide a more qualitative and predictive understanding of glass transition.

The layout of the paper is as follows. We consider an isolated system
consisting of the system of interest surrounded by a very large medium and
follow the consequences of it in the next section. In Sect.
\ref{Marker_Internal_Equilibrium}, we follow the consequence of partial
equilibrium to develop a very general non-equilibrium thermodynamics, which is
then applied to a glass in Sect. \ref{Sec.Local_Equilibrium}. The concept of
the fictive temperature and the Tool-Narayanaswamy phenomenology are
considered in Sect. \ref{Marker_Fictive}, and established on a firm
theoretical ground. The conclusions are given in the last section.

\section{Consequences of the Second Law\label{Marker_Second_Law}}

As said above, we study non-equilibrium systems by proceeding in a general
manner by following the consequences of the second law, which is well
established. As usual, we apply the second law to an isolated system, which we
denote by $\Sigma_{0}$; it\ consists of the system $\Sigma$ of interest (such
as our glass) in a medium denoted by $\widetilde{\Sigma}$ containing it. We
will consider a single component system, which is sufficient for our purpose.
According to the second law, the entropy $S_{0}$ of an isolated system
$\Sigma_{0}$ can never decrease in time \cite{Landau}:%
\begin{equation}
\frac{dS_{0}(t)}{dt}\geq0.\label{Second_Law}%
\end{equation}
What happens inside the isolated system (loss of ergodicity in parts of the
system, chemical reactions, phase changes, etc.) cannot affect the direction
of the inequality, which makes it the most general principle of
non-equilibrium thermodynamics. The law itself imposes no restriction on the
actual rate of entropy change. In general, $S_{0}$ also depends on the number
of particles $N_{0}$, energy $E_{0}$, and volume $V_{0}$ of $\Sigma_{0}$.
Thus, $S_{0}(t)$ used above should be really written as $S_{0}(E_{0}%
,V_{0},N_{0},t)$. However, as the extensive quantities remain constant in time
there is no harm in using the compact form $S_{0}(t)$ during approach to
equilibrium.\ The entropy $S_{0}(t)$ is a \emph{continuous} function of each
of its arguments. The energy, volume and the number of particles of $\Sigma
$\ are denoted by $E$, $V$, and $N,$ respectively, while that of the medium
$\widetilde{\Sigma}$ by $\widetilde{E}$, $\widetilde{V}$, and $\widetilde{N}.$
Obviously,%
\[
E_{0}=E+\widetilde{E},\ \ V_{0}=V+\widetilde{V},\ \ N_{0}=N+\widetilde{N}.
\]
We will assume that $N$ of the system is also fixed, which means that
$\widetilde{N}$ is also fixed. However, the energy and volume of the system
may change with $t$.

When the isolated system is in equilibrium, its entropy $S_{0}(E_{0}%
,V_{0},N_{0},t)$ has reached its maximum and no longer has any \emph{explicit}
time-dependence so that it can be simply written as $S_{0}(E_{0},V_{0},N_{0})$
or $S_{0}$. In this case, different parts of $\Sigma_{0}$\ have the same
temperature $T_{0}$ and pressure $P_{0}$:%
\begin{equation}
\frac{1}{T_{0}}=\frac{\partial S_{0}}{\partial E_{0}},\text{\ \ \ }\frac
{P_{0}}{T_{0}}=\frac{\partial S_{0}}{\partial V_{0}}; \label{Eq_Conds_0}%
\end{equation}
we have defined the temperature by setting the Boltzmann constant
$k_{\text{B}}=1$ in this work. Otherwise, the entropy $S_{0}(t)$
\emph{continuously increases} and the isolated system is said to be not in
equilibrium. The medium is considered to be very large compared to $\Sigma,$
so that its temperature, pressure, etc. are not affected by the system. We
assume $\widetilde{\Sigma}$ to be in internal equilibrium (its different parts
have the same temperature and pressure, but $\widetilde{\Sigma}$ and $\Sigma$
may not be in equilibrium with each other). Thus, its entropy $\widetilde{S}$
no longer has an explicit time dependence, but has an \emph{implicit}
$t$-dependence through the $t$-dependence of $\widetilde{E}$, and
$\widetilde{V}$. The time variation of $S_{0}(t)$ is due to the relaxation
going on inside $\Sigma$ as it is driven towards equilibrium with the medium.

The entropy $S_{0}(t)$ of the isolated system can be written as the sum of the
entropies $S(t)$ of the system and $\widetilde{S}(t)$ of the medium:%
\begin{equation}
S_{0}(E_{0},V_{0},N_{0},t)=S(E,V,N,t)+\widetilde{S}(\widetilde{E}%
,\widetilde{V},\widetilde{N});\label{Total_Entropy}%
\end{equation}
there is no explicit $t$-dependence in $\widetilde{S}(\widetilde{E}%
,\widetilde{V},\widetilde{N})$ due to internal equilibrium. The correction to
this entropy due to the weak stochastic interactions between the system and
the medium has been neglected, which is a common practice \cite{Landau}. We
expand $S_{0}$ in terms of the small quantities of the system \cite{Landau}%
\[
\widetilde{S}(\widetilde{E},\widetilde{V},\widetilde{N})\simeq\widetilde
{S}(E_{0},V_{0},\widetilde{N})-\left.  \left(  \frac{\partial\widetilde{S}%
}{\partial\widetilde{E}}\right)  \right\vert _{E_{0}}E(t)-\left.  \left(
\frac{\partial\widetilde{S}}{\partial\widetilde{V}}\right)  \right\vert
_{V_{0}}V(t).
\]
It follows from the internal equilibrium of $\widetilde{\Sigma}$ that%
\[
\left.  \left(  \frac{\partial\widetilde{S}}{\partial\widetilde{E}}\right)
\right\vert _{E_{0}}=\frac{1}{T_{0}},\ \ \left.  \left(  \frac{\partial
\widetilde{S}}{\partial\widetilde{V}}\right)  \right\vert _{V_{0}}=\frac
{P_{0}}{T_{0}},
\]
and $\widetilde{S}\equiv\widetilde{S}(E_{0},V_{0},\widetilde{N}),$ which is a
constant, is independent of the system. Thus,%
\begin{equation}
S_{0}(t)-\widetilde{S}\simeq S(E,V,N,t)-E(t)/T_{0}-P_{0}V(t)/T_{0}%
.\label{Total_Subtracted_Entropy}%
\end{equation}

Let us introduce
\begin{equation}
G(t)\equiv H(t)-T_{0}S(t),\ H(t)\equiv E(t)+P_{0}V(t),\label{Free_Energies}%
\end{equation}
the time-dependent Gibbs\ free energy and enthalpy of the system $\Sigma$ with
the medium $\widetilde{\Sigma}$ at fixed $T_{0}$ and $P_{0}$. We thus finally
have
\begin{equation}
S_{0}(t)-\widetilde{S}=S(t)-H(t)/T_{0}=-G(t)/T_{0}%
,\label{Gibbs_Free_Energy_Entropy_Relation}%
\end{equation}
so that the behavior (\ref{Second_Law}) of $S_{0}(t)$\ of the isolated system
leads to an very important conclusion about the Gibbs free energy of the
system:
\begin{equation}
\frac{dG(t)}{dt}\leq0.\label{Gibbs_Free_Energy_Variation}%
\end{equation}
The Gibbs free energy $G(t)$ decreases as the system relaxes towards
equilibrium, a result quite well known. If we abruptly cool the system from
some previous temperature such as $T_{\text{g}}$ to a lower temperature, the
Gibbs free energy at the new temperature remains equal to its value at the
previous temperature at time $t=0$. As it relaxes, $G(t)$ continuously
decreases. It cannot increase without violating the second law.

We have given the essential steps in its derivation here not only for the sake
of continuity as some of the intermediate steps will be needed later on, but
also to make some important points, which we now list.

\begin{enumerate}
\item In deriving the above equation (\ref{Gibbs_Free_Energy_Entropy_Relation}%
), no assumption about the system $\Sigma$ has been made. In particular, we
have not assumed any particular aspect of its non-equilibrium nature, such as
a particular form of relaxation (Arrhenius or otherwise), loss of ergodicity, etc.

\item The identification of $S_{0}(t)-\widetilde{S}$ with the Gibbs free
energy $G(t)$ of $\Sigma$ is generally valid under the assumption of the
medium being large compared to $\Sigma$, which can be satisfied as well as we wish.

\item The Gibbs free energy $G(t)$ and the enthalpy $H(t)$ are determined by
the temperature $T_{0}$ and the pressure $P_{0}$ of the large medium.

\item The continuity of $S_{0}(t)$ with respect to all of its arguments that
was mentioned earlier also applies to the Gibbs free energy of the system.

\item The decrease in $G(t)$ must not be violated even when there is a loss of
ergodicity in the system, as is commonly believed to occur in a glass transition.

\item For glasses, we have an additional experimental fact. The enthalpy
remains continuous across the glass transition. The continuity of the enthalpy
then implies that the entropy $S(t)$ also will remain continuous with respect
to all of its arguments, as the system relaxes. This allows us to
differentiate the entropy, which will be required in Sect.
\ref{Marker_Internal_Equilibrium}.
\end{enumerate}

\section{Non-equilibrium Thermodynamics\label{Marker_Internal_Equilibrium}%
\qquad\qquad}

When the equality in (\ref{Second_Law}) occurs, different parts of $\Sigma
_{0}$\ (such as $\Sigma$\ and $\widetilde{\Sigma}$)\ have the same temperature
$T_{0}$ and pressure $P_{0}$.\ Otherwise, they have different temperatures and
pressures, in which case a common assumption made by almost all workers is
that of partial equilibrium (see, for example, Landau and Lifshitz \cite[see
p. 13]{Landau}) when $\Sigma_{0}$ is out of equilibrium; each part is in
\emph{internal equilibrium }(local equilibrium), which then allows us to
define the temperature, pressure, etc. for each part, which may all be
different. In this situation, their entropies have the maximum possible values
for their respective energies and volumes, and the number of particles. As a
result, they have no explicit $t$-dependence [see the equilibrium condition
(\ref{Eq_Conds_0}) above for $\Sigma_{0}$]; their variation in time comes from
the time variation of their energies, volumes, etc. The entropy $S$ of the
system determines its instantaneous temperature $T(t)$ and pressure $P(t)$:%
\begin{equation}
\frac{\partial S}{\partial E}=\frac{1}{T(t)},\ \frac{\partial S}{\partial
V}=\frac{P(t)}{T(t)}. \label{Eq_Conds}%
\end{equation}
These are standard relations for the entropy \cite{Landau}, except that all
quantities except $S$ in the above equations may have an \emph{explicit}
dependence on time $t$ that will make $S$ depend \emph{implicitly} on time.
Accordingly,
\begin{equation}
\frac{\partial S}{\partial t}=0 \label{Eq_Conds_t}%
\end{equation}
under internal equilibrium. Relations like (\ref{Eq_Conds}) along with
(\ref{Eq_Conds_t}) for internal equilibrium are used commonly in
non-equilibrium thermodynamics. For example, we use them to establish that
heat flows from a hot body to a cold body; see for example Sect. 9 in Landau
and Lifshitz \cite{Landau}. The glassy state, in which the fast dynamics has
equilibrated and the slow dynamics is extremely slow, will thus be treated as
a state in internal equilibrium, although it is not in equilibrium (with the
medium).\textbf{ }This observation will be very important when we discuss the
concept of the fictive temperature in Sect. \ref{Marker_Fictive}.

Recognizing that $S(t)$ has no explicit $t$-dependence, see (\ref{Eq_Conds_0}%
), but is a function of $E(t)$ and $V(t)$ ($N$ is kept a constant), we have
for the differential $dS(t)$%
\[
dS(t)=\frac{1}{T(t)}dE(t)+\frac{P(t)}{T(t)}dV(t),
\]
where we have used (\ref{Eq_Conds}) and have allowed the pressure and the
temperature of the system to be different from those of the medium for the
sake of generality. The first law of thermodynamics follows from this
equation:
\begin{equation}
dE(t)=T(t)dS(t)-P(t)dV(t),\label{First_Law}%
\end{equation}
which does not depend on the temperature and pressure of the medium. In this
form, the first law has the standard look with the first term representing the
heat
\begin{equation}
dQ=T(t)dS(t)\label{Heat}%
\end{equation}
added to the system and the second term without the sign denoting the work
\[
dW=P(t)dV(t)
\]
done by the system.

Using $H(t)=E(t)+P_{0}V(t)$ in (\ref{Free_Energies})$,$ we find that
\begin{equation}
dH(t)=T(t)dS(t)+V(t)dP_{0}+[P_{0}-P(t)]dV(t), \label{First_Law_Enthalpy}%
\end{equation}
where the last term appears due to the lack of equilibrium with the medium.
Accordingly, the heat $dQ,$ see (\ref{Heat}), is no longer equal to
$dH(t)$\ at constant pressure $P_{0}$ of the medium:%
\[
dQ(t)=\left.  dH(t)\right\vert _{P_{0}}+[P(t)-P_{0}]dV(t).
\]
The specific heat $C_{P}$\ at constant pressure is given by%
\[
C_{P}(t)\equiv\left(  \frac{\partial H(t)}{\partial T_{0}}\right)  _{P_{0}%
}+[P(t)-P_{0}]\left(  \frac{\partial V(t)}{\partial T_{0}}\right)  _{P_{0}}%
\]
However, obtaining the entropy of the system from the measured values of the
specific heat requires care:%
\[
dS(t)=\frac{C_{P}}{T(t)}dT_{0}\leq\frac{C_{P}}{T_{0}}dT_{0},
\]
which follows from (\ref{Temp_Relaxation}), derived later.

The differential of $G(t)$, see (\ref{Free_Energies}), turns out to be%
\begin{equation}
dG(t)=-S(t)dT_{0}+V(t)dP_{0}+[T(t)-T_{0}]dS(t)+[P_{0}-P(t)]dV(t),
\label{First_Law_Gibbs}%
\end{equation}
Again, the last two terms are corrections to $dG(t)$ due to the
non-equilibrium nature of the system. We observe from (\ref{First_Law_Gibbs})
that%
\begin{align*}
\left(  \frac{\partial G}{\partial T_{0}}\right)  _{P_{0}}  &
=-S(t)+[T(t)-T_{0}]\left(  \frac{\partial S(t)}{\partial T_{0}}\right)
_{P_{0}}-[P(t)-P_{0}]\left(  \frac{\partial V(t)}{\partial T_{0}}\right)
_{P_{0}},\\
\left(  \frac{\partial G}{\partial P_{0}}\right)  _{T_{0}}  &
=V(t)+[T(t)-T_{0}]\left(  \frac{\partial S(t)}{\partial P_{0}}\right)
_{T_{0}}-[P(t)-P_{0}]\left(  \frac{\partial V(t)}{\partial P_{0}}\right)
_{T_{0}}.
\end{align*}
Again, the last two terms in each equation are the correction due to
non-equilibrium nature of the process, and would be absent in an equilibrium process.

One can compare the Gibbs free energy differential in (\ref{First_Law_Gibbs})
with the approach developed by de Donder \cite{Donder} and Prigogine
\cite{Prigogine}. The last two terms in (\ref{First_Law_Gibbs}) denote the
contributions from two different structural order parameter or the degree of
advancement. In the present context, the two parameters are determined by the
instantaneous entropy and volume
\[
\xi_{1}\equiv\frac{S(t)-S(\infty)}{S(0)-S(\infty)},\ \xi_{2}\equiv
\frac{V(t)-V(\infty)}{V(0)-V(\infty)},
\]
and the corresponding affinities are given by%
\[
A_{1}\equiv-[T(t)-T_{0}]\left[  S(0)-S(\infty)\right]  ,\ A_{2}\equiv\lbrack
P(t)-P_{0}]\left[  V(0)-V(\infty)\right]  .
\]
Thus, we have%
\[
\lbrack T(t)-T_{0}]dS(t)-[P(t)-P_{0}]dV(t)\equiv-A_{1}\xi_{1}-A_{2}\xi_{2},
\]
so that each contribution $A_{i}\xi_{i}\leq0,$ as expected from the variation
of $G(t)$ during relaxation at constant $T_{0},P_{0}$. We also note that we
can write the first law as%
\begin{equation}
dE(t)=T_{0}dS(t)-P_{0}dV(t)-A_{1}\xi_{1}-A_{2}\xi_{2}%
,\label{de_Donder_Thermodynamics}%
\end{equation}
as expected from the standard formulation by de Donder.

\section{Relaxation below the Glass transition\label{Sec.Local_Equilibrium}}

We now apply the above formalism to a glass. Above the glass transition
temperature $T_{\text{g}}$ but below the melting temperature, the system is a
supercooled liquid (SCL) as the relaxation time $\tau$\ of $\Sigma$ remains
less than the observation time $\tau_{\text{obs}}$. At $T_{\text{g}}$, they
become identical. Below $T_{\text{g}}$, $\tau$ becomes larger than
$\tau_{\text{obs}}$, and the system turns into a glass. Let us consider the
system in the glassy state. With time, the glass ($\Sigma$) will relax so as
to come to equilibrium (the corresponding SCL state obtained by increasing
$\tau_{\text{obs}}$ to the relaxation time at that temperature) with the
medium if we wait longer than $\tau_{\text{obs}}$. It should be noted again
that, due to the internal equilibrium of the system, there is no explicit
$t$-dependence in $S$ on the right side of (\ref{Total_Entropy}) or
(\ref{Total_Subtracted_Entropy}). Accordingly, the $t$-dependence in $S,H$,
and $G$ is implicit through $E(t),$ and $V(t)$. In the following, the glass is
considered to be formed under \emph{isobaric} conditions so the pressure of
the medium is kept fixed at $P_{0}$ as its temperature is step-wise varied.
Accordingly, its instantaneous pressure $P(t)$ is always equal to $P_{0}$ of
the medium%
\begin{equation}
P(t)=P_{0}, \label{Isobaric_Cond}%
\end{equation}
but its temperature will in general be different than $T_{0}$ and vary in
time, as we will show below. The initial enthalpy $H(0)$ at $T_{0}$ is the
enthalpy of the glass at temperature $T^{\prime}$, and $H(\infty)$ the value
of the SCL enthalpy after complete relaxation at temperature $T_{0}$. It is
experimentally found that the enthalpy decreases with time during an
isothermal relaxation so that
\begin{equation}
H(t=0)>H(t\rightarrow\infty). \label{Enthalpy_Inequality}%
\end{equation}
This decrease is a general property of thermodynamics which follows from the
specific heat being non-negative. We first consider the case when $T^{\prime}$
is the glass transition temperature, so that the system is in the SCL state at
$T^{\prime}$. As $T^{\prime}>T_{0},$ the enthalpy $H(0)$ of SCL at $T^{\prime
}$ must be higher than the enthalpy $H(t\rightarrow\infty)$ of SCL at $T_{0}$.
The same is also true of the volume in many cases, which relaxes to a smaller
value in an isothermal relaxation. However, this property of the volume is not
a thermodynamic requirement. Accordingly, as a general rule%
\begin{equation}
\frac{dH(t)}{dt}<0, \label{Relaxation_Facts}%
\end{equation}
during isothermal structural relaxation in glasses. In the following, we will
only use the above general property of the enthalpy, and not of the volume.
Now, if $T^{\prime}$ is below the glass transition, then from the result just
derived, we conclude that $H(0)$ is even larger than the SCL enthalpy at
$T^{\prime}$. This even strengthens the above inequality
(\ref{Enthalpy_Inequality}).

Let us now turn to the time derivative of the entropy $S_{0}$, which is
changing because the energy and volume of $\Sigma$ are changing with time
\cite{Landau}. Thus,%
\begin{align*}
\frac{dS_{0}(t)}{dt} &  =\frac{dS}{dt}-\frac{1}{T_{0}}\frac{dE(t)}{dt}%
-\frac{P_{0}}{T_{0}}\frac{dV(t)}{dt}\\
&  =\left(  \frac{\partial S}{\partial E}-\frac{1}{T_{0}}\right)  \frac
{dE(t)}{dt}+\left(  \frac{\partial S}{\partial V}-\frac{P_{0}}{T_{0}}\right)
\frac{dV(t)}{dt}\geq0,
\end{align*}
as the relaxation goes on in the system $\Sigma$. It is clear that
\[
\frac{\partial S}{\partial E}\neq\frac{1}{T_{0}},\ \frac{\partial S}{\partial
V}\neq\frac{P_{0}}{T_{0}},
\]
if $dS_{0}/dt>0$. Thus, as long as the relaxation is going on due to the
absence of equilibrium, the two inequalities must hold true. Accordingly, the
derivative $\partial S/\partial E,$ which by definition \ represents the
inverse temperature $1/T(t)$ of the system, see (\ref{Eq_Conds}), must be
\emph{different} from $1/T_{0}$ of the medium:
\[
T(t)\neq T_{0}.
\]
As $\partial S/\partial V=P_{0}/T(t)$, we see immediately that
\begin{equation}
\frac{dS_{0}(t)}{dt}=\left(  \frac{1}{T(t)}-\frac{1}{T_{0}}\right)
\frac{dH(t)}{dt}\geq0.\label{Total_Entropy_Rate}%
\end{equation}
From (\ref{Relaxation_Facts}), we observe $dH(t)/dt<0$ during relaxation in
glasses. Thus, we are forced to conclude that
\begin{equation}
T(t)\geq T_{0},\label{Temp_Relaxation}%
\end{equation}
the equality occurring only when equilibrium has been achieved. The
instantaneous temperature can be measured by using a small "thermometer" so as
not to disturb the internal equilibrium of the glass. Such a measurement will
allow us to explore its variation in time.%
\begin{figure}
[ptb]
\begin{center}
\includegraphics[
trim=1.001315in 7.225579in 2.790484in 0.647003in,
height=2.7968in,
width=4.3837in
]%
{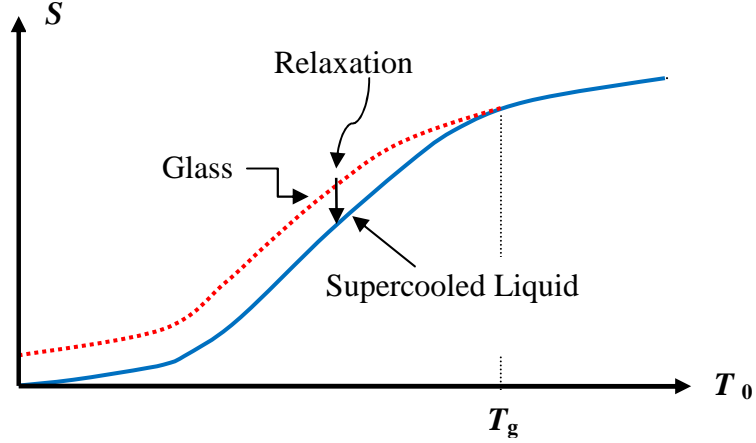}%
\caption{Schematic behavior of the entropy for SCL (blue curve) and GL\ (red
dotted curve). The GL entropy decreases, shown by the downward arrow, as it
isothermally (constant temperature $T_{0}$ of the medium) relaxes towards SCL,
during which its temperature $T(t)$ also decreases towards $T_{0}$. }%
\label{Fig_entropyglass}%
\end{center}
\end{figure}

The above calculation also shows that
\begin{equation}
\frac{dS(t)}{dt}=\frac{1}{T(t)}\frac{dH(t)}{dt}, \label{Entropy_variation1}%
\end{equation}
which is the first term in (\ref{Total_Entropy_Rate}). The equation above can
also be obtained from (\ref{First_Law_Enthalpy}). Using $P(t)=P_{0}$ for an
isobaric process, which is the normal situation in most experiments, we see
that the last term in (\ref{First_Law_Enthalpy}) vanishes. Hence,
\[
dQ=dH(t)=T(t)dS(t)
\]
is valid in all isobaric processes, from which (\ref{Entropy_variation1})
follows immediately.

The relaxation that occurs in the glass originates from its tendency to come
to thermal equilibrium during which its temperature $T(t)$ varies with time;
recall that we are considering a cooling experiment. The relaxation process
results in the lowering of the corresponding Gibbs free energy, as is seen
from (\ref{Gibbs_Free_Energy_Variation}), which is a consequence of the second
law in (\ref{Second_Law}). Accordingly, there are changes in its enthalpy and
entropy, which are in the same direction; see (\ref{Entropy_variation1}). The
lowering of $G(t)$ with time results in not only lowering the enthalpy in a
cooling experiment, as observed experimentally, but also the entropy $S(t)$
during relaxation:%
\[
(dS(t)/dt)\leq0,
\]
as shown in Fig. \ref{Fig_entropyglass}. At constant $P_{0}$ and $T_{0}$, we
see from (\ref{First_Law_Gibbs}) that
\begin{equation}
dG(t)=[T(t)-T_{0}]dS(t),\label{First_Law0}%
\end{equation}
from which it follows that
\[
\frac{dG(t)}{dS(t)}=T(t)-T_{0},
\]
showing that $G(t)$ converges to its equilibrium value more slowly compared to
the convergence of $S(t)$ as $T(t)\rightarrow T_{0},$ i.e. as $t\rightarrow
\infty$ as the above derivative vanishes in this limit.

It is instructive to compare the specific heat of the glass with the specific
heat of the corresponding fully relaxed state obtained as $t\rightarrow\infty
$. Let us assume that at time $t=0,$ we change the temperature of $\Sigma$
form some initial temperature $T^{\prime}$ to $T_{0}$ instantaneously. We
consider the system at $t=t_{\text{obs}}$ and determine its enthalpy. The
specific heat of the glassy sample at this instant is given by%
\[
C_{P,\text{g}}=\lim_{\Delta T\rightarrow0}\frac{H(0)-H(t_{\text{obs}})}{\Delta
T}.
\]
Then, the corresponding specific heat after complete relaxation is given by%
\[
C_{P\text{,relax}}=\lim_{\Delta T\rightarrow0}\frac{H(0)-H(\infty)}{\Delta
T}\geq C_{P,\text{g}}.
\]

\section{The Fictive Temperature and the Tool-Narayanaswamy
Equation\label{Marker_Fictive}}

We now need to turn our attention to the distinction between the fast and slow
degrees of freedom (dof), a characteristic of any glass. Situation similar to
this also occurs in the attainment of thermal equilibrium between the nuclear
spins and their environment during nuclear relaxation \cite{Purcell}, where
the spin-lattice relaxation is extremely slow. The explanation of this kind of
behavior (slow and fast dof) in a wide class of substances lies in internal
molecular motions other than simple vibrations. The fast dof cool down and
equilibrate very fast, while the slow dof take much longer to transfer their
energy and equilibrate because of very weak coupling with the surrounding
medium. Here, we are talking about equilibration with the medium. We will
denote those dof that have equilibrated with the medium at time $t$ by a
subscript "e", and the remaining that are not equilibrated by "n."

We set $t=0$ at the instant the system is abruptly cooled from its equilibrium
state at $T^{\prime}$ to some lower temperature $T_{0}$ of the medium. At this
time, all dof are out of equilibrium with the medium at $T_{0}$. The fast dof
equilibrate within the observation time $t_{\text{obs}},$\ with the slow dof
remaining out of equilibrium \cite{Goldstein}.\ Eventually, all dof come to
equilibrium with the medium. Thus, the number of dof in equilibrium keeps on
increasing with time. Let $D$ denote the total number of the dof in the
system, which is determined by the number of particles $N$ in it; hence, it
remains constant. Let $D_{\text{e}}(t)$ and $D_{\text{n}}(t)$ denote its
partition in equilibrated and non-equilibrated dof, respectively:%
\[
D=D_{\text{e}}(t)+D_{\text{n}}(t);
\]
evidently, they keep varying in time. As said above, the clear distinction
between the two kinds of dof arises because of a very weak coupling between
them and of the slow dof with the medium. The weak coupling allows us to treat
them as almost uncorrelated and \emph{quasi-independent}, which then
immediately leads to the following partition of the entropy, the energy and
the volume into two contributions, one from each kind because of their
quasi-independence mentioned above:
\begin{subequations}
\label{Partitions}%
\begin{align}
S(t) &  =S_{\text{e}}(t)+S_{\text{n}}(t),\label{Partitions_S}\\
E(t) &  =E_{\text{e}}(t)+E_{\text{n}}(t),\label{Partitions_E}\\
V(t) &  =V_{\text{e}}(t)+V_{\text{n}}(t),\label{Partitions_V}%
\end{align}
where the notation is self-evident. It should be noted that\ $S_{\text{e}}%
(t)$\ and $S_{\text{n}}(t)$\ stand for $S_{\text{e}}(E_{\text{e}%
}(t),V_{\text{e}}(t))$ and $S_{\text{n}}(E_{\text{n}}(t),V_{\text{n}}(t)).$
The corrections to each of the partition due to the very weak coupling is
small enough to be neglected.\ 

One should not confuse dof$_{\text{e}}$ with only fast dof. To see this most
clearly, we recall that at $t=0$, none of the dof have equilibrated at the new
temperature $T_{0}$, so that $D_{\text{e}}(t=0)=0$. But all of the same dof
were equilibrated at $T^{\prime}$, implying that dof$_{\text{e}}$ at
$T^{\prime}$ contains fast and slow dof. The same happens as $t\rightarrow
\infty$, in which case $D_{\text{e}}(t)\rightarrow D$ at $T_{0}$ implying that
all dof, fast and slow, have equilibrated. Thus, in general, $D_{\text{e}}(t)$
contains both fast and slow dof. Let us now consider dof$_{\text{n}}$. At
$t=0$, none of the dof have equilibrated at the new temperature $T_{0}$.
Accordingly, $D_{\text{n}}(t=0)=D,$ so that dof$_{\text{n}}$ contains both
fast and slow dof. However, for $t>t_{\text{obs}}$, only (or mostly) the slow
dof remain in $D_{\text{n}}(t)$.

Let us now introduce the following derivatives of the energy partitions%
\end{subequations}
\begin{equation}
x(t)\equiv\frac{dE_{\text{e}}(t)}{dE(t)},\ 1-x(t)\equiv\frac{dE_{\text{n}}%
(t)}{dE(t)}, \label{x_definition}%
\end{equation}
at a given $t$, so that
\begin{equation}
\frac{\partial S_{\text{e}}(t)}{\partial E(t)}=x(t)\frac{\partial S_{\text{e}%
}(t)}{\partial E_{\text{e}}(t)},\ \ \frac{\partial S_{\text{n}}(t)}{\partial
E(t)}=[1-x(t)]\frac{\partial S_{\text{n}}(t)}{\partial E_{\text{n}}(t)}.
\label{x_T_derivatives}%
\end{equation}
The derivatives in the two equations above are at fixed $V_{\text{e}}(t)$\ and
$V_{\text{n}}(t)$, respectively. At $t=0$, $D_{\text{e}}(t=0)=0,$
$E_{\text{e}}(t=0)=0$ and $x(t=0)=0.$ At $t\rightarrow\infty$, $D_{\text{e}%
}(t)\rightarrow D,\ E_{\text{n}}(t)=0$ so that $x(t)=1$. As time goes on, more
and more of the "n" dof equilibrate, thus increasing $D_{\text{e}}(t)$ and
$x(t)$.\qquad

By definition, we have
\[
\frac{\partial S_{\text{e}}(t)}{\partial E_{\text{e}}(t)}=\frac{1}{T_{0}},
\]
which follows from the equilibrium of the dof$_{\text{e}}$ with the medium,
while the dof$_{\text{n}}$ will have a temperature different from this.
Accordingly, we introduce a new temperature $T_{\text{n}}(t)$, defined by the
derivative%
\begin{equation}
\frac{\partial S_{\text{n}}(t)}{\partial E_{\text{n}}(t)}=\frac{1}%
{T_{\text{n}}(t)}. \label{Fictive_Temp}%
\end{equation}
The following identity%
\begin{equation}
\frac{1}{T(t)}=\frac{x(t)}{T_{0}}+\frac{1-x(t)}{T_{\text{n}}(t)}
\label{Narayanaswamy_Decomposition}%
\end{equation}
easily follows from considering $\partial S(t)/\partial E(t)$ and using
(\ref{Partitions_S}) and (\ref{x_T_derivatives}). Initially, $x(0)=0$ so that
$T(0)=$ $T_{\text{n}}(0)=T^{\prime}$, while $T(t)\rightarrow T_{0}$ as
$t\rightarrow\infty$, as expected. This division of the instantaneous
temperature $T(t)$ into $T_{0\text{ }}$and $T_{\text{n}}(t)$ is identical in
form to that suggested by Narayanaswamy \cite{Narayanaswamy}, except that we
have given thermodynamic definitions of the non-linearity parameter $x(t)$ in
(\ref{x_definition}) and of the new temperature $T_{\text{n}}(t)$
(\ref{Fictive_Temp}) in our approach. Both these quantities, being intensive,
can only depend on energy and volume per particle, through which these
quantities gain their implicit $t$-dependence.

Let us now understand the significance of the above analysis. The partition of
the thermodynamic quantities in (\ref{Partitions}) along with the definition
of the fraction $x(t)$ shows that the partition satisfies a lever rule: the
relaxing glass can be \emph{conceptually} (but not physically) thought of as a
"mixture" consisting of two different "components" corresponding to
dof$_{\text{e}}$ and dof$_{\text{n}}$: the former is at temperature $T_{0}$
and has a weight $x(t)$; the latter with a complementary weight $1-x(t)$ is at
a temperature $T_{\text{n}}(t).$ Thinking of a system conceptually as a
"mixture" of two "components" is quite common inn theoretical physics. One
common example is that of a superfluid, which can be thought of as a "mixture"
of a normal viscous "component" and a superfluid "component" \cite[Sect.
23]{Landau2}. In reality, there exist two simultaneous motions \cite{Landau2},
one of which is "normal" and the other one is "superfluid". A similar division
can also be carried out in a superconductor: the total current is a sum of a
"normal: current and a "superconducting current" \cite[Sect. 44]{Landau2}.

The division of the dof envision above is no different from these divisions in
a superfluid or a superconductor. However, because of the non-equilibrium
nature of the system, there is an important difference here compared to a
superfluid or a superconductor. The e-component is in equilibrium (with the
medium), but the n-component is only in internal equilibrium. While the
significance of the former as a SCL "component" at $T_{0},P_{0}$
(dof=$D_{\text{e}}$) is obvious, the significance of the latter requires
clarification. At $t=0$, $T_{\text{n}}(t)$ represents the temperature
$T^{\prime}$ of the equilibrated SCL (dof=$D$) from which the current glass is
obtained by cooling. At this time, the entropy $S_{\text{n}}(t=0)$\ of initial
state of the glass at $T_{0}$ is equal to the entropy $S_{\text{SCL}%
}(T^{\prime})$\ of the equilibrated SCL (dof=$D$) at the previous temperature
$T^{\prime}$. The latter has the energy and volume $E_{\text{SCL}}(T^{\prime
})=E_{\text{n}}(t=0)$ and volume $V_{\text{SCL}}(T^{\prime})=V_{\text{n}%
}(t=0).$ At any later time $t>0$, $T_{\text{n}}(t)$ represents the temperature
associated with the energy $E_{\text{n}}(t)$ and volume $V_{\text{n}}(t)$ of
the non-equilibrated "component" of the glass and has a weight $1-x(t)$. This
"component," being in internal equilibrium, can be identified as a
\emph{fictive} SCL [dof=$D_{\text{n}}(t)$]\ at temperature $T(t)<T^{\prime}$
of energy $E_{\text{SCL}}=E_{\text{n}}(t)$ and volume $V_{\text{SCL}%
}=V_{\text{n}}(t).$ In other words, the relaxing glass at any time $t$ can be
considered as consisting of two SCL "components," one at temperature $T_{0}$
[dof=$D_{\text{e}}(t)$] and the other one [dof=$D_{\text{n}}(t)$] at
temperature $T_{\text{n}}(t)$. The temperature $T\equiv T_{\text{n}}(t)$
\emph{uniquely} determines the energy $E_{\text{SCL}}(T)\equiv E_{\text{n}%
}(t)$ and volume $V_{\text{SCL}}(T)\equiv V_{\text{n}}(t)$ of the
corresponding \emph{fictive} SCL [dof=$D_{\text{n}}(t)$].

As the fictive liquid at $T\equiv T_{\text{n}}(t)$ contains only (or mostly)
the slow dof, it does not yet really represent a SCL associated with the
system at $T\equiv T_{\text{n}}(t)$, as the former lacks dof$_{\text{e}}$,
while the latter contains all dof. This does not pose any problem as the
missing dof$_{\text{e}}$ at $T\equiv T_{\text{n}}(t)$ are in equilibrium not
only with the dof$_{\text{n}}$ at $T\equiv T_{\text{n}}(t)$, the fictive SCL
mentioned above, but also with the medium at $T\equiv T_{\text{n}}(t)$. Thus,
one can consider "adding" these missing dof$_{\text{e}}$ (dof=$D_{\text{e}}$)
to the fictive liquid, which now represents the equilibrated SCL (dof=$D$) at
$T\equiv T_{\text{n}}(t)$. This SCL is not the same as the glass with its
fictive $T_{\text{n}}(t)$, as the latter has its dof$_{\text{e}}$ at $T_{0}$
while the SCL has all of its dof at $T_{\text{n}}(t)$. However, all of their
thermodynamic properties associated with dof$_{\text{n}}$ must be the same, as
their entropy function is the same for both liquids. Similarly, the SCL
"component" at $T_{0},P_{0}$ (dof=$D_{\text{e}}$) should also be
"supplemented" by the missing dof$_{\text{n}}$ to give rise to the
equilibrated SCL at $T_{0},P_{0}$ (dof=$D$).

We are now in a position to decide which of the temperatures $T(t)$ and
$T_{\text{n}}(t)$ qualifies as the \emph{fictive temperature}. This
temperature is supposed to characterize the non-equilibrium aspect of the
system. As $T(t)$ contains information about both kinds of dof, it is not the
appropriate temperature to be identified as the fictive temperature, even
though it depends on $t$. The temperature $T_{\text{n}}(t)$, on the other
hand, depends only on non-equilibrated dof$_{\text{n}}$, \ and should be
identified as the fictive temperature of the relaxing glass at time $t$. This
temperature is not the internal temperature of the glass at this time, but
represents the equilibrium temperature of the corresponding SCL at $T\equiv
T_{\text{n}}(t)$ as noted above.

As first pointed out by Littleton \cite{Littleton} and Lillie \cite{Lillie},
and discussed by several authors, see for example \cite{Cox,Scherer90,Hodge},
the viscosity keeps changing with time during relaxation. Thus, if one uses an
Arrhenius form for the viscosity, it must depend not on $T_{0}$, but on
$T(t)$; it is the instantaneous temperature that characterizes the
instantaneous state of the glass. Thus, the Arrhenius form for the viscosity,
which is usually taken to be proportional to the relaxation time, must be
expressed as
\begin{equation}
\eta(t)=\eta_{0}\exp\left[  \frac{B}{T(t)}\right]  =\eta_{0}\exp\left[
B\left(  \frac{x(t)}{T_{0}}+\frac{1-x(t)}{T_{\text{n}}(t)}\right)  \right]
,\label{Narayanaswamy_Relaxation}%
\end{equation}
the form conventionally identified as the phenomenological Tool-Narayanaswamy
form. Here, $\eta_{0}$ and $B$ are some parameters of the system and may
depend on $T_{0},P_{0}$ and also weakly on time $t$. Our derivation above
justifies this form on a solid theoretical ground.

One can carry out a similar analysis with decomposing the volume; see
(\ref{Partitions_V}). However, we do not obtain any new result as
$P(t)=P_{0}.$ To see this, we proceed exactly as above but use the volumes
instead of the energies. Introducing the parameter $x_{v}(t)$ defined by
\[
x_{v}(t)\equiv\frac{\partial V_{\text{e}}(t)}{\partial V(t)},
\]
at fixed $E_{\text{e}}(t)$, which may be different from $x(t),$ and using
\[
\frac{\partial S_{\text{e}}(t)}{\partial V_{\text{e}}(t)}=\frac{P_{0}}{T_{0}%
},\ \frac{\partial S_{\text{n}}(t)}{\partial V_{\text{n}}(t)}=\frac{P_{0}%
}{T_{\text{n,}v}(t)},\
\]
at fixed $E_{\text{e}}(t)$\ and $E_{\text{n}}(t)$, respectively, along with
$\partial S(t)/\partial V(t)=P_{0}/T(t),$\ see (\ref{Eq_Conds}) and
(\ref{Isobaric_Cond}), we find the following decomposition of the inverse
instantaneous temperature
\[
\frac{1}{T(t)}=\frac{x_{v}(t)}{T_{0}}+\frac{1-x_{v}(t)}{T_{\text{n,}v}(t)}.
\]
Now, the new fictive temperature $T_{\text{n,}v}(t)$ represents the
temperature $T=$ $T_{\text{n,}v}(t)$ of corresponding fictive SCL
[dof=$D_{\text{n}}(t)$] with energy and volume $E_{\text{SCL}}(T)=E_{\text{n}%
}(t)$ and volume $V_{\text{SCL}}(T)=V_{\text{n}}(t).$ This fictive liquid is
the same as noted above as far as the energy and volume are concerned.
However, as SCL is an equilibrated state, the specification of energy and
volume \emph{uniquely} determines the temperature, which from the earlier
analysis was seen to be exactly $T_{\text{n}}(t).$ Thus, we conclude
\[
T_{\text{n,}v}(t)\equiv T_{\text{n}}(t).
\]
In order for the above decomposition to be consistent with
(\ref{Narayanaswamy_Decomposition}) at all times, we must ensure that%
\[
x_{v}(t)=x(t).
\]
In other words, our definition of the fictive temperature gives the same value
whether we consider the energy relaxation or the volume relaxation.

It is highly likely that the slow relaxation consists of many different
relaxation modes, which we index by $j=1,2,\cdots.$ However, there does not
seem to be any strong argument to suggest that all these different relaxation
modes are almost decoupled as was the case for the fast and slow relaxations
\cite{Goldstein} studied above. In that case, it is not possible to partition
the thermodynamic quantities such as the entropy, etc. associated with
dof$_{\text{n}}$ as a sum over these different modes. Despite this, let us
follow the consequences of such an assumption. We express $D_{n}(t)$ as a sum
over $j$%
\[
D_{n}(t)\equiv\sum_{j}D_{\text{n}}^{(j)}(t),
\]
where the notation is quite transparent. We similarly express all of the
n-quantities in (\ref{Partitions}) as a sum over $j$. We can similarly express%
\[
1-x(t)\equiv\sum_{j}y_{j},
\]
where
\[
y_{j}\equiv\frac{dE_{\text{n}}^{(j)}(t)}{dE(t)}.
\]
We can now introduce a fictive temperature for each $j$-th n-dof%
\[
\frac{\partial S_{\text{n}}^{(j)}(t)}{\partial E_{\text{n}}^{(j)}(t)}=\frac
{1}{T_{\text{n}}^{(j)}(t)},
\]
such that
\[
\frac{1-x(t)}{T_{\text{n}}(t)}\equiv\sum_{j}\frac{y_{j}}{T_{\text{n}}%
^{(j)}(t)},
\]
a decomposition also described by Narayanaswamy \cite{Narayanaswamy}. However,
because of the above equality, the presence of more than one kind of
relaxation modes does not change the earlier decomposition
(\ref{Narayanaswamy_Decomposition}). In other words, no new insight is gained
by such an assumption. One can introduce an equilibrated fictive SCL at each
of the fictive temperatures as above. We will not stop here to do so as it is straightforward.

\section{Conclusions}

We have developed a non-equilibrium thermodynamics to study systems away from
equilibrium. The approach is quite general and is not limited to systems close
to equilibrium. Assuming internal equilibrium, a common practice in the field,
we find the correction to the differential free energies that are consistent
with de Donder-Prigogine approach to non-equilibrium thermodynamics; see for
example (\ref{de_Donder_Thermodynamics}). Even though, we have mainly
discussed supercooled liquids, the approach does not require the presence of a
melting transition and an equilibrium crystal for its application. Thus, it
should also be applicable to other glassy systems such as spin glasses, where
there is no equilibrium crystal as the true equilibrium state. The only
requirement is that enough time has passed after the system has been disturbed
so that the instantaneous temperature, pressure, etc. can be defined via
(\ref{Eq_Conds}) for the system even if they are changing with time. In other
words, there is partial equilibrium in the isolated system. We then apply this
thermodynamics to study glasses and clarify the concept of the fictive
temperature $T_{\text{n}}(t)$ widely used in the study of glasses by
identifying it as a thermodynamic quantity; see (\ref{Fictive_Temp}). Our
analysis shows that the fictive temperature has the same value even if we
change the relaxing quantity. This temperature is not identical to but is
related to the instantaneous temperature $T(t)$ in a glass; see
(\ref{Narayanaswamy_Decomposition}). We use this relationship to establish the
Tool-Narayanaswamy equation (\ref{Narayanaswamy_Relaxation}) for the
relaxation time on a solid theoretical ground. This form does not change even
if we have more than one kind of slow relaxation; the latter results in many
different fictive temperatures, one for each kind of slow relaxation.

We should finally contrast our approach with other approaches available in the
literature. We can use (\ref{Partitions_S}) to express $dS(t)=dS_{\text{e}%
}(t)+dS_{\text{n}}(t).$ However, each term is still multiplied by $T(t),$
implying that $dS_{\text{e}}(t)$ or $dS_{\text{n}}(t)$ are not multiplied by
their respective temperatures $T$ or $T_{\text{n}}(t)$. Thus, one cannot
consider $S_{\text{e}}(t)$ and $S_{\text{n}}(t)$ as separate in the first law
(\ref{First_Law}) or in (\ref{First_Law0}). This should be contrasted with the
approach developed in \cite{Nieuwenhuizen}, where the entropy is divided into
fast and slow dof; see also \cite{Garden}. More recently, M\"{o}ller et al
\cite{Moller} have used the approach of de Donder to study glasses by
considering a single structural order parameter; however, the concept of the
fictive temperature was not analyzed. A recent approach by Wolynes
\cite[(b)]{Wolynes} provides a local description of the relaxation in an
inhomogeneous mosaic form, but the interest is in the dynamics, whereas our
focus is not on any particular dynamics.

\end{document}